\documentclass[aps,prl,twocolumn,showpacs,amsmath,amssymb,floatfix,superscriptaddress]{revtex4-2}

\usepackage{CJK}
\usepackage{graphicx}
\usepackage{hyperref}
\usepackage{color}
\usepackage[yyyymmdd,hhmmss]{datetime}
\usepackage[normalem]{ulem}
\usepackage{todonotes}
\usepackage{multirow}
\usepackage{float}
\usepackage{array, blindtext}
\usepackage{adjustbox}
\usepackage{makecell}
\usepackage{lipsum}
\usepackage{bm}
\usepackage{amsmath}

\begin{document}\textit{}
\begin{CJK*}{UTF8}{gbsn}
\title{Weak Magnetic Sensing via Floquet Driving in an Active Cavity Magnon Coupled System}
\author{Fan Yang (杨帆)}
\affiliation{School of Physics, Shandong University, 27 Shandanan Road, Jinan, 250100 China} 
\author{Xudong Wang (王旭东)}
\email{xudongwang01@mail.sdu.edu.cn}
\affiliation{School of Physics, Shandong University, 27 Shandanan Road, Jinan, 250100 China}
\author{Lijun Yan (闫丽君)}
\affiliation{School of Physics, Shandong University, 27 Shandanan Road, Jinan, 250100 China} 
\author{Yue Zhao (赵越)}
\affiliation{School of Physics, Shandong University, 27 Shandanan Road, Jinan, 250100 China} 
\author{Jinwei Rao (饶金威)}
\affiliation{School of Physics, Shandong University, 27 Shandanan Road, Jinan, 250100 China}
\author{Lihui Bai (柏利慧)}
\email{lhbai@sdu.edu.cn}
\affiliation{School of Physics, Shandong University, 27 Shandanan Road, Jinan, 250100 China}
\author{Shishen Yan (颜世申)}
\affiliation{School of Physics, Shandong University, 27 Shandanan Road, Jinan, 250100 China}

\date{\today/\currenttime}

\begin{abstract}
While significant advancements have been made in weak magnetic field detection, conventional high-sensitivity techniques are often limited by requirements for cryogenic operation or bulky setups. In this work, we develop a sensitive alternating magnetic field sensor based on a coupled system of an active microwave cavity and yttrium iron garnet (YIG), with the components implemented on printed circuit boards (PCBs). By introducing electrically tunable gain to compensate for cavity losses, we substantially improve both the quality factor and the signal intensity. Under the coupled system, Floquet modulation is induced by the alternating magnetic field, allowing for weak field detection by driving a specific hybrid mode and measuring the resulting Floquet sidebands. This miniaturized device operates at room temperature, achieving a detection limit of $121 \text{ pT}/\sqrt{\text{Hz}}$.\\


\end{abstract}

\maketitle

Weak magnetic field detection has remained a focal point of research as a vital tool for exploring fundamental physical phenomena and biological signals. Traditional Superconducting Quantum Interference Devices\cite{jaklevic1964quantum,greenberg1998application,kleiner2004superconducting,granata2016nano}, with their exceptional sensitivity, have played a pivotal role in ultra-low-field nuclear magnetic resonance detection\cite{jiang2013study,clarke2007squid}. Spin-exchange relaxation-free atomic magnetometers have successfully bypassed conventional sensitivity limits and are widely employed in the medical field due to their superior noise suppression capabilities\cite{kominis2003subfemtotesla,savukov2005nmr,savukov2005tunable,wang2025zero,li2023miniaturized,savukov2007detection,wang2024magnetic}.
Concurrently, diamond-based nitrogen-vacancy center magnetometers\cite{wolf2015subpicotesla,wang2022picotesla,fescenko2020diamond,barry2016optical}, utilizing optomagnetic detection, have pushed the spatial resolution of weak field sensing to the nanometer scale, offering an unprecedented observational window.
However, these technologies still face practical challenges, including cryogenic requirements and large physical footprints.
Overcoming these limitations to achieve high-sensitivity, room-temperature, and compact sensing remains a critical objective in current research.

In recent years, cavity magnon coupled systems have witnessed rapid development\cite{PhysRevLett.113.156401,bai2015spin,yang2020unconventional,wang2018bistability,mi2025bifurcation}, providing a well-suited platform for room-temperature and miniaturized weak magnetic field detection\cite{rao2021interferometric,wu2022observation,wang2025speech,crescini2021phase}.
However, traditional planar resonators are hindered by intrinsic dissipations such as Ohmic losses, which limit their quality factors (Q) and constrain their applications in precision measurement.
Recent studies have introduced gain into cavity magnon coupled systems to construct non-Hermitian gain-compensated architectures by leveraging external energy sources to actively counteract internal losses. These systems provide a foundation for high-Q coupling at room temperature\cite{gui2025broadband,zhang2024van,kim2024low,rao2023meterscale,yao2023coherent,wang2025single,zhang2025gain,yan2026manipulating}.
Concurrently, time-domain modulation based on Floquet theory has introduced a new dimension for signal amplification and control\cite{yang2023theory,wang2021floquet,xu2020floquet,oka2019floquet,jiang2022floquet}.
Through alternating magnetic field modulation, the system generates sideband signals in the frequency domain, enabling an amplified response to external alternating fields.
Implementing Floquet modulation within an active cavity magnon system would theoretically yield even narrower linewidths and stronger responses to external fields, boosting detection sensitivity.
Despite its potential for sensitive magnetic sensing, Floquet engineering in active cavity magnon coupled systems remains unexplored.

In this work, we demonstrate a high-sensitivity alternating magnetic field sensing platform based on an active cavity magnon coupled system.
By using tunable gain to compensate for cavity losses, we significantly enhance the quality factor and boost the signal intensity.
In the strong coupled system, we drive one hybrid mode while the target alternating field induces Floquet modulation.
By monitoring the resulting sidebands at the other mode, we achieve linear detection of weak magnetic fields. 
Experimental results show that the system achieves a magnetic sensitivity of $121\text{ pT}/\sqrt{\text{Hz}}$ at 80 MHz.
This work highlights the significant potential of integrating gain mechanisms with Floquet engineering for advanced magnetic sensing applications.

\begin{figure}
    \centering
    \includegraphics[width=1\linewidth]{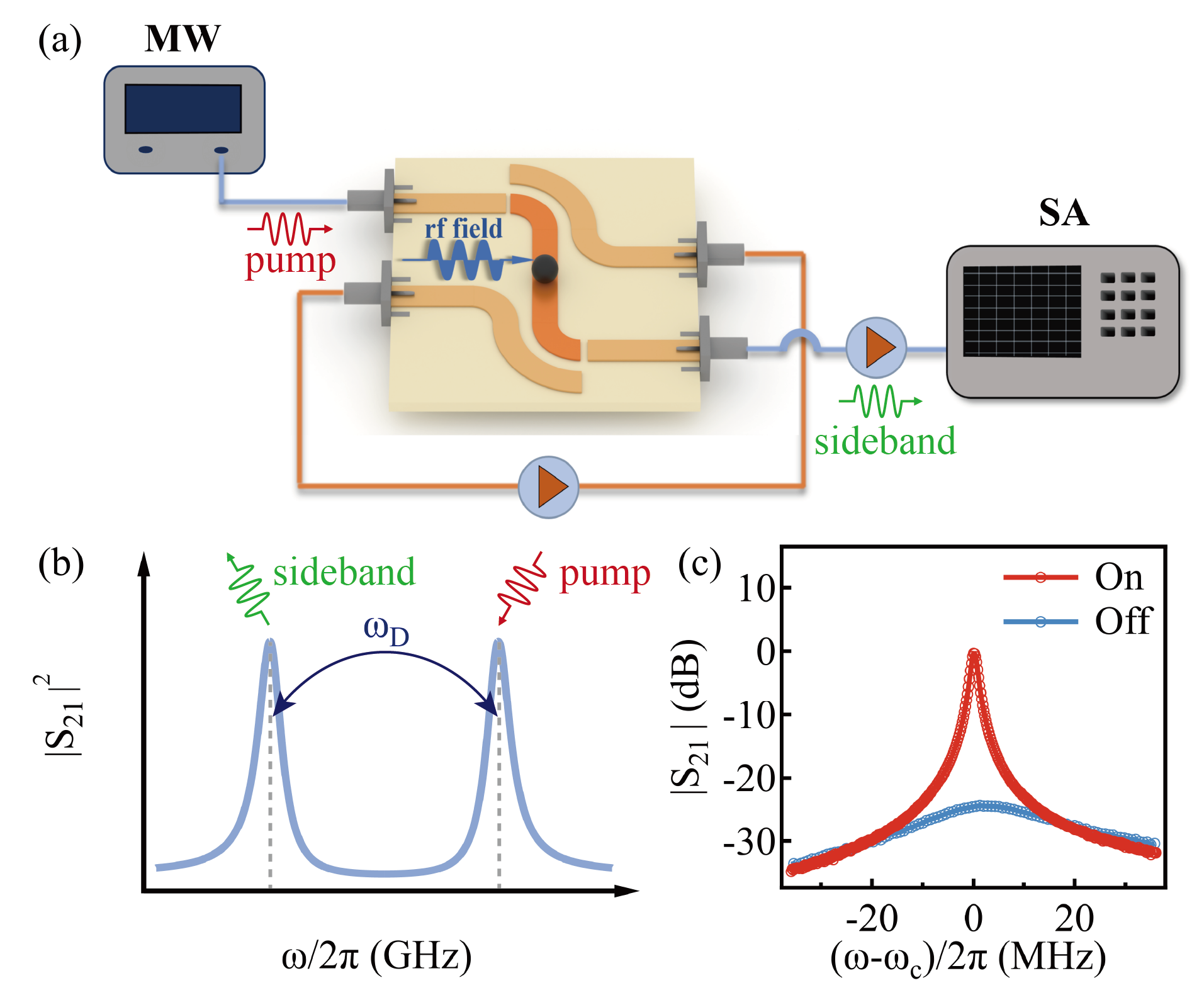}
    \caption{(a) Schematic of the experimental setup. A microwave pump drives the coupled system, while an alternating magnetic field is applied to the YIG sphere. The resulting output sideband signals serve as a linearly amplified representation of the alternating field for readout.
    (b) Working principle schematic. The light blue line denotes the transmission spectrum of the coupled system, while the dark blue line indicates the frequency of the alternating magnetic field. The pump is applied to the higher frequency mode, and the resulting sideband signals are detected at the lower frequency mode.
    (c) Transmission spectrum of the cavity during gain off and on states.}
    \label{fig.1}
\end{figure}

The cavity magnon coupled system, as shown in Fig. \ref{fig.1}(a), consists of a planar resonant cavity and a Yttrium Iron Garnet (YIG) sphere with a diameter of 1.2 mm.
The highlighted orange region in the schematic represents the cavity, with the cavity mode frequency given by $\omega_c/2\pi = 4.2$ GHz, determined by the length of the resonant cavity. 
The signal trace (highlighted in light orange) on the printed circuit board (PCB) is connected to an amplifier via a discrete wire, establishing a gain circuit.
The static magnetic field $B$ is used to adjust the precession frequency of the YIG sphere, thereby altering the coupling position between the magnon mode and the cavity mode. 
Additionally, an inductive coil is wound around the vicinity of the YIG sphere to generate an alternating magnetic field $b$ that is parallel to the static field $B$.
The coupled system is composed of two hybrid modes. At zero detuning  ($\omega_m=\omega_c$), the frequency splitting between these modes is equal to twice the coupling strength of the system, defined as $2g$, as shown in Fig. \ref{fig.1}(b).
By applying an alternating magnetic field $b$ at frequency $\omega_D$ and driving one of the hybrid modes (e.g., $\omega_+$) with a microwave field, the resonance frequency undergoes periodic modulation. 
This process leads to the generation of sidebands at $\omega_+ \pm n\omega_D$. 
Based on this phenomenon, the intensity of the sideband signal can be used to characterize the magnitude of the alternating magnetic field.
Due to the frequency-selective nature of the cavity magnon system, the sideband intensity peaks with the condition $\omega_+ \pm n\omega_D = \omega_-$ satisfied, where our subsequent measurements are conducted.
To further amplify the sideband signal, we introduce gain to compensate for the system's damping, which significantly narrows the mode linewidth.
As illustrated in Fig. \ref{fig.1}(c), the cavity mode without gain exhibits a low-amplitude, broad profile.
In contrast, the application of gain results in a much sharper resonance and a substantial increase in signal intensity, with the quality factor of the cavity enhanced from 121 to 4600.

 \begin{figure}
     \centering
     \includegraphics[width=1\linewidth]{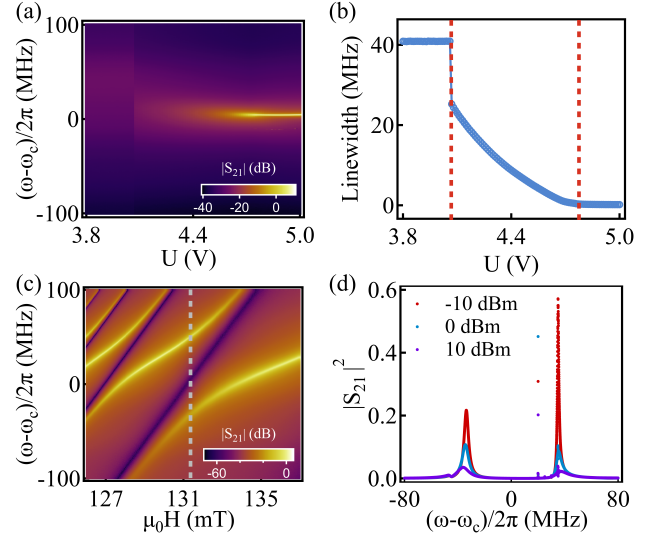}
     \caption{(a) Transmission spectra of the active cavity at various bias voltages.
     (b) Cavity mode linewidth extracted from (a) as a function of bias voltage.
     (c) Transmission spectra of the cavity magnon coupled system. The grey dashed line indicates the position of zero detuning.
     (d) Transmission spectrum of the coupled system at zero detuning under pump powers of -10, 0, and 10 dBm.}
     \label{fig.2}
 \end{figure}

The amplifier acts directly on the microwave photon modes within the cavity to provide single-photon gain. Upon introducing this gain, the complex frequency of the cavity mode is expressed as $\widetilde{\omega}_c = \omega_c - i\kappa_c = \omega_c - i[\beta + \kappa + \gamma |c|^2 - G]$.
Here, $\kappa_c$ denotes the total decay rate of the cavity mode, while $\beta$ and $\kappa$ represent the intrinsic damping rate and the external damping rate.
$G$ corresponds to the single-photon gain rate.
The term $\gamma |c|^2$ accounts for the nonlinear damping, which is proportional to the photon density $n_c = |c|^2$ and represents additional dissipation induced by nonlinear effects.
As the external bias voltage increases, the single-photon gain rate $G$ rises, causing a corresponding decrease in the total damping $\kappa_c$.
Experimentally, after the gain voltage reaches the amplifier turn-on voltage of 4.06 V, the linewidth of the cavity mode gradually narrows, and its intensity gradually increases as the voltage continues to rise (Fig. \ref{fig.2}(a)).
The linewidth of the cavity mode was extracted using the Lorentz formula and plotted in Fig. \ref{fig.2}(b).
Before the turn-on voltage, the linewidth remains constant as the gain polariton is not yet excited. As the gain voltage continues to increase, the linewidth gradually decreases.
The continued rise in voltage drives an increase in photon number, bringing the nonlinear damping term $\gamma \left | c \right | ^{2}$ into competition with the optical gain.
Beyond a gain voltage of 4.73 V, $\kappa_c$ reaches its minimum value and plateaus.
This indicates that the nonlinear damping and the gain have reached a dynamic balance, which in turn leads to stabilization of the spectral linewidth.
Coupling the active cavity mode to the magnon mode yields an anti-crossing dispersion relation, as presented in Fig. \ref{fig.2}(c).
At the magnetic field indicated by the dashed line, the system operates under a zero-detuning condition, where the pump microwave resonantly excites the higher-frequency hybrid mode.
However, the pump power affects the photon number in the cavity, which in turn influences the gain regime.
As illustrated in Fig. \ref{fig.2}(d), varying the pump power modifies the gain state of the system.
At a low power of -10 dBm, the limited number of pump-induced photons allows the system to remain in a self-sustained state.
Conversely, at 10 dBm, the high photon density enhances significant nonlinear damping, which increases the total dissipation and consequently broadens the linewidth and suppresses the signal intensity.
Therefore, we performed all subsequent measurements under an optimized pump power of 0 dBm.

\begin{figure}
    \centering    \includegraphics[width=1\linewidth]{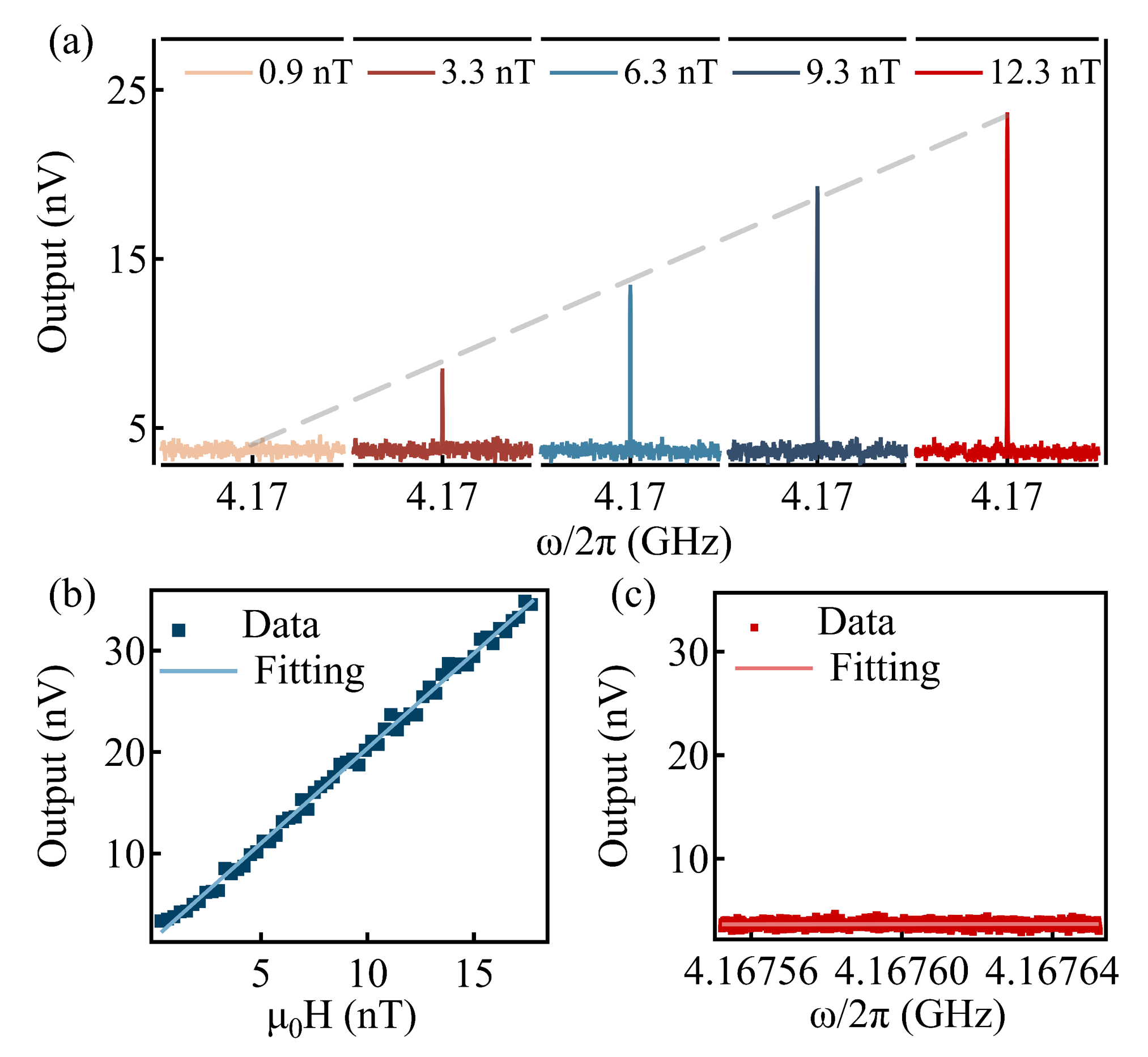}
    \caption{(a) Sideband signal spectra under different alternating magnetic fields. (b) Linear dependence of the sideband signal amplitude on the alternating magnetic field strength, with a slope $K$ = 1.81 V/T and an intercept $T$ = 1.49 nV. (c) The average noise level of the measurement system, measured to be 3.68 nV.}
    \label{fig.3}
\end{figure}

The alternating magnetic field $b$ applied to the YIG sphere modulates the spin precession, resulting in a periodic modulation of the spin resonance frequency. The following equations can describe the dynamics of the system:
\begin{equation}
    i \frac{d}{dt} 
    \begin{bmatrix}
c(t)\\m(t)
\end{bmatrix}
=  \begin{bmatrix}
        \widetilde{\omega}_c & g \\
        g & \widetilde{\omega}_m + \Omega_D \cos(\omega_D t)
    \end{bmatrix} 
        \begin{bmatrix}
c(t)\\m(t)
\end{bmatrix}
+ \begin{bmatrix}
    \Omega_{p}e^{-i\omega_{p}t}\\0
\end{bmatrix}
    \label{eq.H}
\end{equation}

 where\  $c(t)$\ and\ $m(t)$\ represents the state vector consisting of the cavity mode and the magnon mode.
 $\Omega_D$\ is the amplitude of the alternating magnetic field\ $b$.  $\Omega_p$ denotes the intensity of the pump microwave, and $\omega_p$ represents its frequency.

 In the limit of weak alternating modulation where $\Omega_D$ approaches zero, the high-order sidebands of the coupled system are negligibly weak. By employing Floquet theory to calculate the first-order sideband intensity, the output sideband intensity of the system is derived as follows(See supplementary materials for details):
\begin{widetext}
 \begin{equation}
    A_{\pm 1} = \frac{i \sqrt{\kappa} g^2 \Omega_{p} (\Omega_D/2)}{ \left[ \omega_D(\omega_D \pm 2g) - \kappa_c \alpha_m - i(\pm \omega_D - g)(\kappa_c + \alpha_m) \right] \left[ ig(\kappa_c + \alpha_m) - \kappa_c \alpha_m \right] }
    \label{eq.A}
\end{equation}
\end{widetext}
The derived expression indicates that the output sideband intensity is directly proportional to the alternating field strength.
To validate this relationship, we further conducted experimental measurements for quantitative verification.

The alternating magnetic field $b$ applied to the YIG sphere modulates the spin precession, leading to a periodic modulation of the spin resonance frequency.
By driving the higher-frequency mode at $\omega_2 = $4.25 GHz, a sideband signal emerges at $\omega_1 = $4.17 GHz under an alternating field at $\omega_b = $80 MHz.
The sideband intensity increases with the magnitude of the alternating field $b$, as illustrated by the spectra for five different field strengths in Fig. \ref{fig.3}(a).
Specifically, the signal is buried in the noise floor at $b = $0.9 nT, whereas it becomes clearly distinguishable with a high signal-to-noise ratio at $b = $3.3 nT.
Fig. \ref{fig.3}(b) shows the linear fit of the sideband intensity versus the alternating field strength, yielding a slope of k = 1.81 V/T, which is in excellent agreement with the theoretical prediction in Eq.\ref{eq.A}.
Furthermore, the noise power spectrum measured by a spectrum analyzer (SA) is presented in Fig. \ref{fig.3}(c).
At a 100 Hz bandwidth, the measured noise floor corresponds to an equivalent magnetic field of 1.21 nT, yielding a sensitivity of $121 \text{ pT}/\sqrt{\text{Hz}}$ for the 80 MHz alternating field.

In conclusion, we have designed and demonstrated a scheme for alternating magnetic field detection based on an active cavity magnon coupled system. By introducing electrically tunable gain, the damping of the cavity mode is effectively suppressed, leading to a significant improvement in the quality factor and signal strength of the resonator. Under strong coupled conditions, detection of an 80 MHz alternating magnetic field is achieved via the Floquet sideband response, with an estimated sensitivity of $121 \text{ pT}/\sqrt{\text{Hz}}$.
These results demonstrate the efficacy of integrating active cavity with Floquet modulation in enhancing magnetic field detection and pave the way for the development of room-temperature, high-sensitivity, and compact magnetometers.

\section{Acknowledgements}
Project supported by the National Key Research and Development Program of China (Grant No.2023YFA1406604), the National Natural Science Foundation of China (Grant Nos. 12274260, 12522405, 12474120, and 12504141), the Shandong Provincial Natural Science Foundation, China (Grant Nos. ZR2024YQ001, ZR2025QC1475), the Taishan Young Scholar Program of Shandong Province (Grant No. tsqn202507065).

\end{CJK*}  

\bibliographystyle{iopart-num.bst}
\bibliography{reference}

\end{document}